\documentstyle[prl,aps,twocolumn,epsfig]{revtex}

\begin{document}
\draft
\preprint{}
\title{\bf Nonextensivity of
 the cyclic Lattice Lotka Volterra model}
\author{G. A. Tsekouras$^{1,2}$, A. Provata$^{1}$ and C. Tsallis$^3$ \thanks{ aprovata@limnos.chem.demokritos.gr, tsallis@cbpf.br}
 }
\address {$^1$ Institute of Physical
Chemistry, National Research Center ``Demokritos'', 15310 Athens, Greece \\
$^2$ Department of Physics, University of Athens, 10679 Athens, Greece\\
$^3$ Centro Brasileiro de Pesquisas Fisicas, Rua Xavier Sigaud 150, 22290-180
Rio de Janeiro - RJ, Brazil}
\date{\today}
\maketitle
\begin{abstract}
We numerically show that the Lattice Lotka-Volterra model, when realized on a 
square lattice support, gives rise to a {\it finite} production, per unit time, of the nonextensive entropy $S_q= \frac{1- \sum_ip_i^q}{q-1}$ $(S_1=-\sum_i p_i \ln p_i)$. 
This finiteness only occurs for $q=0.5$ for the $d=2$ growth mode (growing droplet), and for $q=0$ 
for the $d=1$ one (growing stripe). This strong evidence of nonextensivity is consistent with the spontaneous emergence 
of local domains of identical particles with fractal boundaries and competing interactions. 
Such direct evidence is for the first time exhibited for a 
many-body system which, at the mean field level, is conservative. 

\end{abstract}


{\bf Pacs Numbers:} 5.40.-a, 05.10.Ln, 05.65.+b, 82.40.Np

\vspace{1.0cm}


Many natural and artificial systems are known today to be hardly, or not at all, tractable within 
Boltzmann-Gibbs statistical mechanics, hence the usual thermodynamics. Such is the case of
 systems which include long-range interactions or long-range microscopic (or mesoscopic) memory, 
or other sources of (multi)fractality. Phenomena where many spatial and/or temporal scales are involved 
typically exhibit power laws. The celebrated Boltzmann-Gibbs (BG) entropy $S_{BG}=-\sum_i p_i \ln p_i$ appears
 to be inadequate for handling the thermostatistics associated with such situations. 
This is due to the fact that the corresponding stationary states do  {\it not} emerge through ergodic dynamics. 
An ubiquitous class of the above anomalous systems have nonlinear dynamics 
which generate  {\it weak} chaos, in the sense that the sensitivity to the initial conditions is 
less-than-exponential in time. Such situations quite naturally accomodate with an entropy which generalizes 
the BG one, namely
\begin{equation}
S_q= \frac{1- \sum_ip_i^q}{q-1} \;\;\;(q \in {\cal R}; S_1=S_{BG}). 
\end{equation}
For independent systems $A$ and $B$ (i.e., such that $p_{ij}^{A+B}=p_i^A \, p_j^B$), 
this entropy satisfies $S_q(A+B)=S_q(A) + S_q(B)+(1-q)S_q(A)S_q(B)$. It is due to this property 
of nonextensivity that the thermostatistical formalism based on Eq. (1) is usually referred to as 
nonextensive statistical mechanics \cite{tsallis} (see \cite{review} for recent reviews). 
This theory has received many applications in areas such as self-gravitating polytropes \cite{plastino}, 
electron-positron annihilation \cite{bediaga}, turbulence \cite{turbulence}, motion of {\it Hydra viridissima} 
\cite{arpita}, anomalous diffusion \cite{alemany,difusao}, classical \cite{lyra} 
and quantum \cite{weinstein} chaos, long-range-interacting many-body Hamiltonians \cite{rapisarda}, option pricing \cite{lisa}, particular biological processes involving a large number of
degrees of freedom \cite{tamarit,bezerianos}, among others.    

 The lattice Lotka-Volterra (LLV) model \cite{LLV} is known to well mimic simple chemical reactions, 
predator-prey systems, and other biological and echological phenomena. It has recently been studied with cyclic interactions
 amongst three species, as a modification to the original Lotka-Volterra model\cite{LV}.
The LLV has exhibited spatial  clustering and fractality, even when implemented on low
dimensional lattices \cite{tsekprov}. Since fractality is a distinct sign of nonextensivity, it is natural to 
pose the question whether the LLV model is indeed consistent with the nonextensive premises.
The aim of the present paper is to verify that this model exhibits a strong and direct evidence of having $q \ne 1$.
In particular
we study the LLV model with various initial conditions, namely at a) the domain formation mode, b) the 
nucleus (or droplet) growth mode, c) the stripe growth mode and d) the roughening
mode. Modes (b) and (c) enable, as we shall see, the direct calculation of $q$. Several analytical or numerical 
calculations of $q$ exist already in the literature, but this is the first time such evidence is directly provided, 
through the time evolution of $S_q$ itself, on a  
many-body  system which is conservative at the Mean Field (MF) level.

The LLV model is a minimal complexity model, with  MF  conservative dynamics
which can be directly 
implemented on lattice and involves only
two reactive species $X_1$ and $X_2$
(adsorbed on a lattice support) and the empty
sites of the support $S$. All reactive steps are bimolecular and the
reaction occurs via hard core interactions. Schematically, the LLV
model has the following form\cite{LLV}:
\begin{mathletters}
\setcounter{equation}{2}
\begin{eqnarray}
X_1 \,+ \,X_2 \quad & &\mathop{\rightarrow}^{k_s} \quad 2X_2 \\
X_2 \,+ \,S   \quad & &\mathop{\rightarrow}^{k_1} \quad 2S \\
S \,+ \,X_1   \quad & &\mathop{\rightarrow}^{k_2} \quad 2X_1
\end{eqnarray}
\end{mathletters}
In particular, a particle $X_1$ adsorbed on a lattice site changes its state
into $X_2$ when it is found in the neighborhood of another $X_2$ particle.
This step (2a) is an autocatalytic reactive step. A particle $X_2$ desorbs 
leaving an empty site $S$, if in the
neighborhood another empty site $S$ is found. This step (2b) is a cooperative
desorption step.
Finally, a particle
$X_1$ can be adsorbed on an empty lattice site $S$ if in the neighborhood another
$X_1$ particle is found. This step (2c) is a cooperative adsorption step.

\par We now recall briefly some of the mean field (MF) and lattice properties of the LLV,
which have been studied in detail in previous works\cite{LLV,tsekprov}.
\par In the MF approximation the LLV model, Eqs. (1), can be described  via the
kinetic, rate  equations:
\begin{mathletters}
\setcounter{equation}{3}
\begin{eqnarray}
{dx_1\over dt}&=&x_1(-k_sx_2+k_2s) \\
{dx_2\over dt}&=&x_2(k_sx_1-k_1s) \\
{ds\over dt}  &=&s(-k_2x_1+k_1x_2)
\end{eqnarray}
\end{mathletters}
where $x_1$, $x_2$ and $s$ correspond to the mean coverage of the 
lattice with particles $X_1$, $X_2$ and empty sites $S$ respectively.
In Eqs. (3),  the mean coverages satisfy identically the conservation condition
$x_1+x_2+s =C$, 
where $C$ is a constant which can be chosen equal to unity,
 corresponding to interpreting $x_1$, $x_2$ and $s$ as fractions of the overall
lattice respectively occupied by $X_1$ particles, $X_2$ particles or being empty.
Using $C=1$ it is possible to eliminate one
of the three variables,  say $s=1-x_1-x_2$,  and, to reduce system (3) to two
equations.
This reduced system admits four steady state solutions, three of which are trivial, 
and one non-trivial\cite{LLV}:
\begin{mathletters}
\setcounter{equation}{4}
\begin{eqnarray}
x_{1s} &=& 0,\, x_{2s}=0 \quad(\rm {empty\, lattice}) \\
x_{1s} &=& 1,\, x_{2s}=0 \quad(\rm {lattice\, poisoned\, by\,} X_1) \\
x_{1s} &=& 0,\, x_{2s}=1 \quad(\rm {lattice\, poisoned\, by\,} X_2) \\
x_{1s} &=& {k_1\over{k_1+k_2+k_s}},\, x_{2s}={k_2\over{k_1+k_2+k_s}}
\end{eqnarray}
\end{mathletters}

A linear stability analysis shows that the trivial states are saddle points
while the nontrivial  one is a center 
compatible with an additional constant of motion $C'=x_1^{k_1}
x_2^{k_2}(1-x_1-x_2)^{k_s}$\cite{picard} at the MF level.
Fig. 1a depicts the temporal evolution of the system for typical values for $(k_1,k_2,k_s)$ and 
initial conditions. The black solid line represents the concentration of
$X_1$ and the dashed line the concentration of $X_2$. The motion is periodic
but non-harmonic. The amplitude of the periodic motion, for given parameter
values, depends solely on the initial conditions\cite{LLV,tsekprov}.
At this level of description the system size does not enter into
the calculations since the MF approximation involves only average
concentrations.
\par To mesoscopically describe the system on a lattice,
many details enter: lattice size and geometry,
number of nearest neighbors (coordination number),
interaction range, etc.
To realize the square lattice LLV we adopt a typical Monte Carlo (MC) 
algorithm (details in \cite{LLV,tsekprov}), namely (1) At every microscopic step
one lattice site is randomly chosen; (2) One of the nearest neighbors is also selected randomly; 
(3) If the original chosen site is $X_1$ ($X_2$) and the selected neighbor
is $X_2$ ($S$) then the chosen site changes to $X_2$ ($S$) with probability 
$k_s$ ($k_1$); If the original chosen site is
 $S$ and the selected neighbor
is $X_1$ then the chosen site changes to $X_1$ with probability $k_2$; 
Otherwise the system remains as it is; (4) Return to step 1.

In the MC procedure the unit of time is chosen as $1/N$,
where $N$ is the total number of lattice sites (occupied and empty). 
For example, for
square lattice, $N=L^2$, where $L$ is the linear size of the lattice.
With this choice of micro-time, in one MC step all lattice
sites are, on the average, scanned once. 
In Fig. 1b typical  behavior of the temporal evolution of the MC concentrations
is shown. In particular, the concentrations of $X_1$
is depicted on the full lattice of size $L\times L = 2^8\times 2^8$ 
(solid line)
and on a sublattice of size $l\times l = 2^5\times 2^5$ (dotted line).  Periodic
boundary conditions are used in all simulations. It is clear
that while on the sublattice the concentrations show oscillatory behavior
with added noise,
 on the entire lattice the oscillations   shrink. 
Fig. 2a gives the typical evolution of a system starting from
random initial conditions (Fig. 2a (t=0 MC)). As time increases 
the system develops local domains and each domain behaves as a local
oscillator with specific characteristic frequency. Because
the various domains have different phases, globally, no oscillation
are observed, in contrast with the MF predictions\cite{LLV}.
Moreover, it has been shown\cite{tsekprov} that the different 
species organize in local domains which
present competing interactions and they have fractal boundaries.
In this
figure and hereafter the $X_1$ particles are depicted in grey color,
the $X_2$ in black and the empty sites in white.
 The initial condition was a homogeneous infinite lattice
with equal concentrations
of $X_1, X_2$ particles and empty sites $S$.  Periodic boundary
conditions are used. 
The fractal properties of the spatial structures can be used
to measure the size of the local oscillators\cite{tsekprov} and point
out to a nonextensive formalism for the calculation of its entropy.

To describe the temporal evolution of the entropy with respect
to one of the species, {\it e.g.} $X_1$, we start from a given configuration, with specific initial conditions on lattice
and let the system evolve according to the MC algorithm. 
The choice of the particular species does
not play any role in the entropy calculations, since the model
is cyclic and all the species are equivalent.
As time increases the system passes through various configurations
which we  record at regular temporal intervals. Let us call $C(t)={
\{ C_{ij}(t)\}, i=1,...L; j=1,...L}$ the specific configuration of the
lattice at time $t$, while $C_{ij}(t)$ denotes the state of site $(i,j)$
at time $t$ and
\begin{eqnarray}
 C_{ij} = \left\{ \begin{array}{ll}
                1 & \mbox{if site $(i,j)$ is occupied by $X_1$}\\ \nonumber
               -1 & \mbox{if site $(i,j)$ is occupied by $X_2$}\\ \label{eq4}
                0 & \mbox{if site $(i,j)$ is occupied by $S$}. \nonumber
              \end{array}             
       \right. 
\end{eqnarray}
Within each configuration we introduce a set of M non-overlapping windows
$\{ W_i\} , (i=1, ...M)$, of size $l\times l$ which cover completely the
lattice. The number of windows is $M=n^2=(L/l)^2$. Consequently
 $C(t) = \bigcup_{i=1}^M W_i$. 
Let us denote with $p_i$ the probability that window $i$ is occupied 
by particles $X_1$. If $n_1(i,t)$ is the number of particles $X_1$
inside window $i$ at time $t$ and $n_1(t)$ is the total number of 
$X_1$ particles on the lattice, then
$p_i(t)={{n_1(i,t)} / {n_1(t)}}$.
This probability set into Eq. (1) provides $S_q(t)$. For short times $S_q(t)$ scales as a nonlinear
function of the time, while for large times depends non-linearly on the
system size. This non-linear dependence on the system size is the basic
indication of non-extensivity. The various values of $q$ highlight characteristics
on different length scales in the system. 
As an example, rare events are characterized by low values of $p_i$. For
$q < 1$, the term $p_i^q$ takes relatively large values and gives important
contribution to the function $S_q$. In contrast, 
if $q > 1$, then $p_i^q << p_i$
and the contribution of rare events is negligible.
\par It is well known that 
scaling behavior is proper to systems which present fractality. 
Especially in monofractals only one level of scaling is detected while in
multifractal structures the different scales grow with different
power laws. The $S_q$ entropy is then the appropriate measure of
complexity because it addresses the complexity in different length
scales by appropriate tuning of the $q$ value.
We study $S_q$ with 
different initial conditions. Depending on the degree of 
organization of the initial state, the entropy may increase going to a more disordered
state or decrease going to a more ordered state.

\par Consider first the case of the "domain growth mode",
where initially the system contains
particles $X_1$, $X_2$ and empty sites $S$ randomly
distributed and  with equal probability, as in Fig 2a (t=0MC). 
The initial state is
statistically uncorrelated. This is the state of maximum entropy.
It is easy to calculate $S_q$ for this state as a function
of the window size. Assume that the $X_1$ particles are  the information
carrying sites, while the $X_2$ and $S$ are the medium. This assumption
does not reduce the system complexity since all three species are
equivalent for cyclic reactions. If the three species have the same
concentrations  then the number of $X_1$ particles will be
on the average equal to $x_1(t=0)=L^2/3$.
If we divide the lattice in 
non-overlapping windows of size $l \times l$ then the average
number of $X_1$ particles in a box is $l^2/3$ while the number of
windows is $n^2=L^2/l^2$. The initial values of $p_i$ are:
\begin{eqnarray}
p_i={{l^2/3}\over{L^2/3}}=\left( {l\over L}\right) ^2 \;(\forall i)\;,
\label{eq8} 
\end{eqnarray}
hence $S_q(0)
=[\left( {L / l}\right) ^{2(1-q)}-1]/ (1-q)$. 
As time increases, the various species organize in domains with fractal
boundaries and we call this mode ``domain formation mode''.
The entropy gradually decreases since 
the system evolves from a random
to a more organized state: See Fig. 2a. In Fig. 2b the corresponding
temporal evolution of $S_q$ is shown for
different values of $q$.  
$S_q$ initially undergoes a few oscillations,
while the system organizes in domains of homologous (identical) species, and then
stabilizes into a lower entropy state.

\par To calculate the entropy production rate at the nucleus growth mode, we 
start with a fully organized state consisting only of particles $S$ and we
include a nucleation droplet of infinitesimal radius $r$ placed on
the lattice. The droplet contains particles $X_1$, $X_2$ and $S$ homogeneously
and randomly distributed within the droplet area. As time increases the
droplet grows forming spontaneously several rings of particles $X_1$, $X_2$
and $S$ sequentially.  The widths of the rings shrink with the distance from
the pure $S$ region and when their width becomes zero the typical LLV 
fractal pattern
appears in the middle as can be seen in Fig. 3a. This type of spreading is called the ``nucleus growth mode'' because an initially small
droplet grows in size and finally covers the entire system. This 2-dimensiona ($d=2$) growth leads to a reorganization of the species
which at the beginning were randomly distributed within the infinitesimal
droplet, while they eventually present fractal patterns as in
Fig. 2a. As time increases this typical
pattern will cover the entire lattice and $S_q$ attain the values calculated in the previous case (see Fig. 2b).
\par As seen in Fig. 3b, only the case $q=0.5$ shows a {\it linear} increase
with time during the entropy production duration; behavior is sub-linear for $q > 0.5$ and superlinear 
for $q < 0.5$. In Fig. 3c we see $S_{0.5}$ 
for various system sizes.
They all start linearly, coincide during the entropy production
period, and saturate at different values, in accordance with Eq. (5). The $q$-value 
 does not change with variations of the lattice
size $L$, of the window size $l$, and of the initial
concentrations of reactants within the original droplet.
Note the difference in ordinate scale between Figs. 2b and 3b.
If we inspect closer the $L=500$ steady state of Fig. 3b, the entropy lines
for $t > 150$ present fluctuations similar to the ones in Fig. 2b. An interesting data collapse is shown in Fig. 3d. 
In Fig. 4a,b the entropy of the LLV model, at the ``stripe
growth mode'' is shown. This is a 1-dimensional growth ($d=1$) mode.
The initial state of the system consists of a stripe
of randomly distributed $X_1$, $X_2$ and $S$ embedded in a lattice 
containing only $S$ particles otherwise (Fig. 4a).
 The entropy features at this mode are similar to the ``nucleus growth mode''
but now only $q=0$ produces the linear behavior, hence $q$ depends on $d$.
\par To explore the entropy production due to surface roughening (or 
``roughening mode''), we investigate the case of an interface separating two stripes of
identical  particles. The setup
of the system at the initial state is as follows: On a $500 \times 500$ square lattice 
we create one stripe of size $S \times L = 50 \times 500$
consisting only by $X_1$ particles followed by a stripe of the same size
but consisting only of $X_2$ particles, while the rest of the lattice is
covered by $S$, see Fig. 5a (t=0MC). In the Figs. 5a,
corresponding to times 0 MC, 20 MC, 60 MC and 100 MC respectively, the
originally linear interfaces roughen and the stripes are deformed. The
process is dynamical and all interfaces move to the right with the
same average velocity. While the
size of the stripes is on the average kept constant, fluctuations make
them vary significantly. In fact, after sufficiently long times (depending
on the width $S$ and the size $L$ of the stripes) 
the stripes will mix and the typical fractal patterns of Fig. 2 will reappear.
The entropy increase due to the roughening of the interfaces is shown
in Fig. 5b. After an initial increase due to roughening the entropy
remains constant with statistical fluctuations around the steady state.
This mode is clearly not adequate for extracting the physical value of $q$.

\par In the current study the nonextensive entropic properties of the
LLV model are examined. The special value of $q$ which produces a linear increase of $S_q(t)$ depends on the dimensionality $d$ of the growth: $q=1-1/d\;(d=1,2)$. 
These non-trivial values of $q$ are in accordance with the appearance of fractal spatial structures observed 
in earlier studies for the same system; also, one expects $q=1$ in the $d \to \infty$ limit. Further studies (e.g., sensitivity to the initial conditions, multifractal function $f(\alpha)$, entropy relaxation, aging) are welcome. 

One of us (CT) acknowledges warm hospitality at, as well as financial support by, the Demokritos Center and the Physics Department of the University of Athens.

\begin{figure}
\centerline{\hbox{\psfig{figure=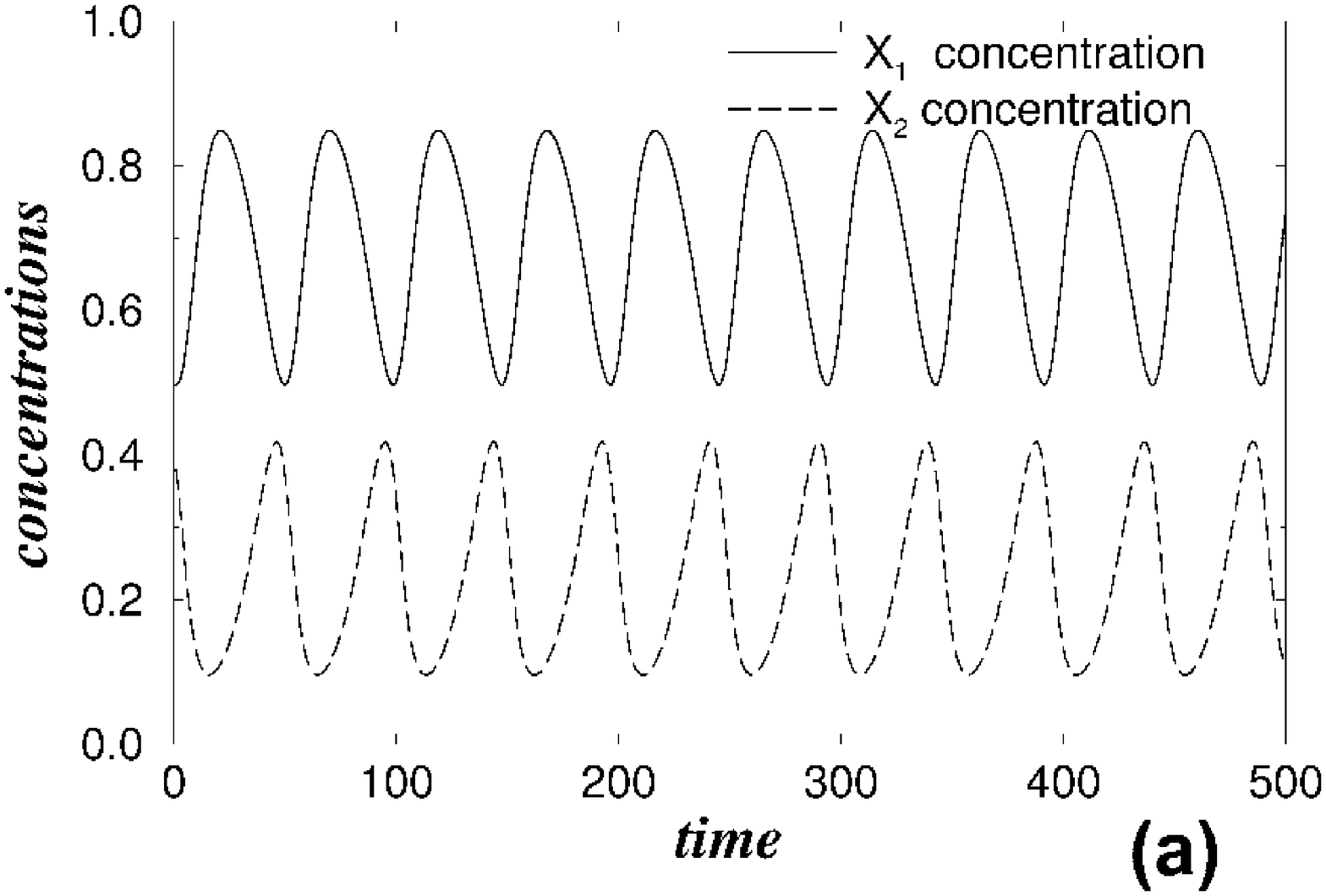,width=3.0cm,height=7.0cm,angle=-90}}}
\centerline{\hbox{\psfig{figure=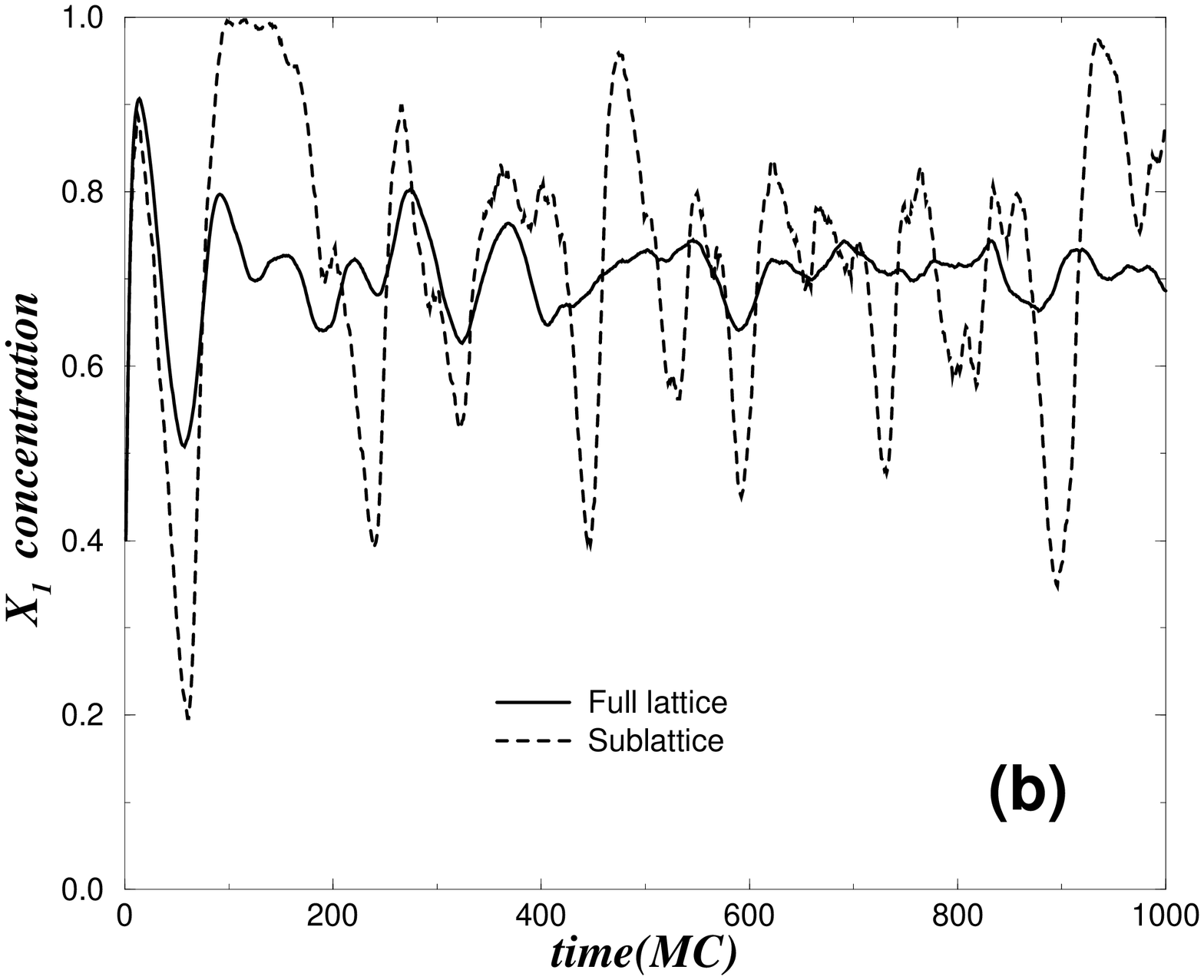,width=3.0cm,height=7.0cm,angle=-90}}}
\label{fig:llv}
\caption{(a) The Mean Field Approximation concentrations $x_1(t)$ and $x_2(t)$ 
 for $(x_1(0), x_2(0),s(0))$ $=(0.5, 0.4,0.1)$, and
  $(k_1, k_2,k_s)$ $=(0.9,0.3,0.1)$.
(b) Monte-Carlo simulations; the solid (dotted) line corresponds to  
$x_1(t)$  over the full lattice (sub-lattice).}
\end{figure}

\begin{figure}[tp]
\centerline{\hbox{\psfig{figure=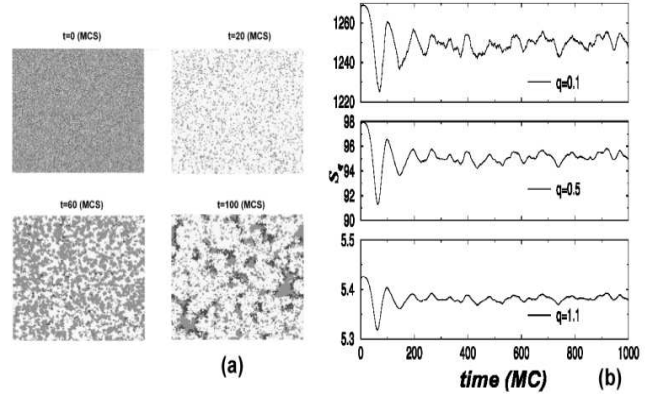,width=9.0cm,height=5.7cm}}}
\caption{(a)Four different snapshots during the evolution of LLV
 for random uniform initial
conditions on a $L=500$ square lattice  at the ``domain formation mode'';
 $(k_1,k_2,k_s)=(0.9.0.3,0.1)$. (b)$S_q(t)$ with $l=10$. }
\end{figure}

\begin{figure}[tp]
\centerline{\hbox{\psfig{figure=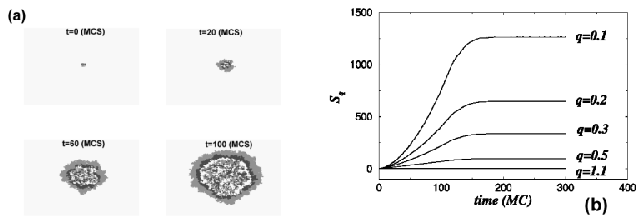,width=9.0cm,height=4.50cm}}}
\vskip 0.1cm
\centerline{\hbox{\psfig{figure=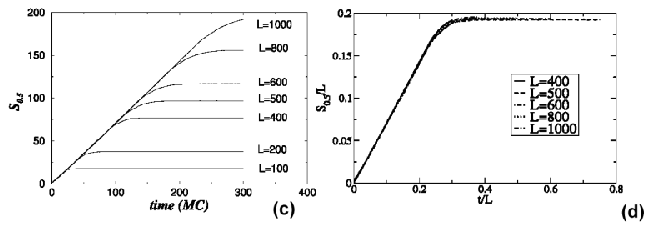,width=9.0cm,height=4.50cm}}}
\label{fig:droplet}
\caption{(a) Four different snapshots during the evolution of a system covered
initially by $S$ with one small mixed droplet. 
The system linear size is $L=500$ while the droplet size is $l=8$;  
$(k_1,k_2,k_s)=(1.0,1.0,1.0)$. (b)$S_q(t)$. (c)$S_{0.5}(t)$ for various lattice sizes.
(d) Collapse of the Fig. 3c data.}
\end{figure}

\begin{figure}[tp]
\centerline{\hbox{\psfig{figure=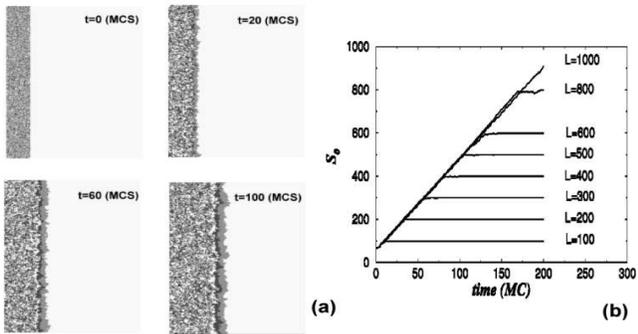,width=9.0cm,height=4.5cm}}}
\label{fig:stripe}
\caption{(a) Four snapshots of the LLV model at the ``stripe growth'' mode; 
$(k_1,k_2,k_s)$ $=(1.0,1.0,1.0)$. (b)$S_0(t)$ for $500 \times L$ lattices.}
\end{figure}

\begin{figure}[tp]
\centerline{\hbox{\psfig{figure=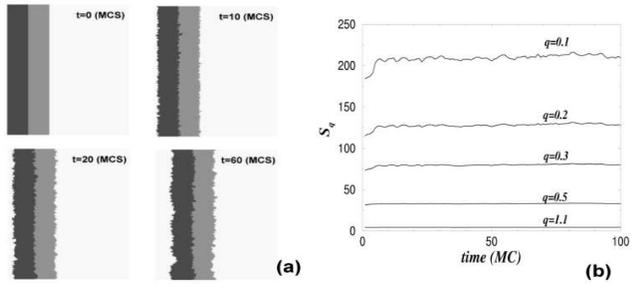,width=9.0cm,height=4.0cm}}}
\label{fig:stripe2}
\caption{(a)Four different snapshots during the evolution of a $L=500$ lattice 
containing initially two stripes of size $50\times 500$; 
$(k_1,k_2,k_s)$ $=(1.0,1.0,1.0)$. (b)$S_q(t)$.}
\end{figure}
\end{document}